\documentclass{elsarticle}
\usepackage[utf8]{inputenc}
\usepackage[T1]{fontenc}
\usepackage[centertags]{amsmath}
\usepackage{amsfonts}
\usepackage{amssymb}
\usepackage{amsthm}
\usepackage{natbib}
\usepackage{caption}
\usepackage{subcaption}
\usepackage{float}
\usepackage{booktabs}
\usepackage{longtable}
\usepackage{xcolor}
\usepackage{hyperref}
\usepackage{algorithm}
\usepackage{algorithmicx}
\usepackage{algpseudocode}

\begin{document}

\begin{frontmatter}

\title{Long-Term Tail Modeling in Survival Analysis via Extended Generalized Pareto Distributions\tnoteref{t1}}

\tnotetext[t1]{Eduardo Janotti is supported by the São Paulo Research Foundation (FAPESP), Brazil, under Grant Nos.\ 2025/10342-3 and 2025/22925-3. This work is funded by national funds through FCT – Fundação para a Ciência e a Tecnologia, I.P., under CEAUL Research Unit, UID/00006/2025, DOI: https://doi.org/10.54499/UID/00006/2025, and by the European Union – NextGenerationEU through the project UID/PRR/00006/2025, DOI: https://doi.org/10.54499/UID/PRR/00006/2025.}

\author[1,3]{Eduardo Janotti\corref{cor1}%
}
\ead{eduardojanotti@ime.usp.br}

\author[2,3]{Lígia Henriques-Rodrigues}
\ead{lcarodrigues@ciencias.ulisboa.pt}

\author[1]{Antonio Carlos Pedroso de Lima}
\ead{acarlos@ime.usp.br}

\cortext[cor1]{Corresponding author}

\affiliation[1]{organization={Departamento de Estatística, Instituto de Matemática e Estatística, Universidade de São Paulo},
city={São Paulo},
country={Brazil}}
\affiliation[2]{organization={Departamento de Ciências Matemáticas, Faculdade de Ciências, Universidade de Lisboa},
city={Lisboa},
country={Portugal}}
\affiliation[3]{organization={CEAUL - Centro de Estatística e Aplicações, Faculdade de Ciências, Universidade de Lisboa},
city={Lisboa},
country={Portugal}}

\begin{abstract}
We propose a class of extended generalized Pareto models for right-censored survival data, with particular emphasis on tail inference and long-term extrapolation. Our framework integrates extreme value theory and survival analysis, combining a generalized Pareto distribution with a flexible perturbation distribution on the unit interval. We consider three perturbation specifications: a parametric Beta model, a Bernstein polynomial estimator, and a histogram-based estimator. To facilitate direct comparison, all three models are fitted using a unified iterative procedure adapted to right censoring through Kaplan–Meier-based pseudo-observations. A Monte Carlo study evaluates finite-sample performance across different tail indices, censoring levels, sample sizes, and model complexities. The results reveal a trade-off between flexibility and stability: the Beta specification generally performs best for tail-index estimation, whereas the histogram estimator performs particularly well for scale estimation under low censoring. The Bernstein estimator shows intermediate performance and greater sensitivity to sample size and censoring. Applications to bladder cancer recurrence and heart-failure survival data show that models with very similar in-sample fits can nevertheless produce markedly different tail-index estimates and long-term extrapolations. These findings emphasize the importance of perturbation specification when extended generalized Pareto models are used for survival extrapolation under censoring.
\end{abstract}


\end{frontmatter}

\section{Introduction}

Many applied problems require the analysis of the time-to-events outcomes, such as mortality or disease recurrence in biostatistics, system breakdown in reliability engineering, or credit default in finance. These problems are typically studied within the framework of survival analysis, which is designed for addressing these questions.

A key complication in this setting is censoring, where the time-to-event is only partially observed for some individuals. As a consequence, standard likelihood-based inference must be adapted in order to accommodate incomplete data, giving rise to widely used techniques such as the Kaplan--Meier estimator \citep{0a032d9f-e977-3fbb-ab72-070bdac732b0} and the Cox proportional hazards model \citep{6c51df5f-27d3-3ed4-9311-1f6af36a828c}.

While these methods are well suited for describing the distribution of observed follow-up times, they provide limited support for extrapolation beyond the observed time horizon. In particular, the Kaplan--Meier estimator becomes flat after the largest observed event time, offering no information about the far-right tail of the survival function. Similarly, semi-parametric regression models generally leave the baseline distribution unspecified, which limits their ability to extrapolate in a principled way.

A commonly used alternative for long-term survival modelling is the class of cure rate models, which assume that a positive fraction of individuals will never experience the event of interest. These models, including mixture \citep{Maller1997} and promotion-time formulations \citep{Yakovlev1996}, interpret the cure fraction as permanent immunity or long-term survival. However, in many practical settings this assumption is difficult to justify, as all those long-term individuals may remain at risk and the apparent plateau may simply reflect limited follow-up rather than non-susceptibility to the event of interest.

When the scientific goal involves long-term risk assessment or the study of rare late events, it becomes necessary to explicitly model the tail of the survival distribution. Tail behaviour plays a key role in a variety of applications, including long-term remission studies, reliability analysis, and epidemiological risk evaluation, where extrapolation beyond observed data is essential.

Extreme Value Theory (EVT) offers a probabilistic foundation for the description of the behaviour of distribution tails and for the performance of extrapolation beyond the range of observed data \citep{coles2001introduction}. By focusing on rare and late events rather than typical central outcomes, EVT provides natural tools to model rare and late-occurring events.

However, incorporating EVT into survival analysis is not straightforward. Classical EVT techniques are typically {developed} under the assumption of complete data and rely on block maxima or threshold exceedances. In contrast, survival data are subject to censoring, which may reduce the effective amount of extreme information available. In addition, standard tail assumptions may not always align with survival contexts, where both light-tailed and heavy-tailed behaviours may arise depending on the application.

Bridging EVT and survival analysis thus requires methods that simultaneously account for censoring and allow for flexible yet principled tail extrapolation. This work is motivated by this need and proposes models that embed extreme value ideas within a survival framework. The aim is to capture long-term behaviour without relying on restrictive cure assumptions while accommodating right censoring.

The main contributions of this work can be summarized as follows. First, we adapt the extended generalized Pareto (EGP) framework to right-censored survival data through a unified iterative estimation procedure. Second, we introduce and compare parametric Beta, Bernstein polynomial, and histogram-based specifications for the perturbation distribution under censoring. Third, we investigate their finite-sample properties through simulation and illustrate their implications for long-term extrapolation using two real-data applications.

The remainder of the paper is organized as follows. Section 2 reviews the EGP framework and establishes the tail-preservation condition used throughout the paper. Section 3 introduces the censored EGP models and estimation procedure. Section 4 presents the simulation study, Section 5 gives the real-data applications, and Section 6 concludes.

\section{EGP framework and tail preservation}

The EGP class provides a flexible framework for modelling heavy–tailed data while retaining the asymptotic properties of the generalized Pareto distribution (GPD). Building on the class introduced by \cite{PAPASTATHOPOULOS2013131}, \cite{CoolArticle} proposed a construction that modifies the bulk of the GPD without altering its tail behaviour. This approach offers several practical advantages: (i) all observations contribute to inference, avoiding information loss caused by threshold exceedance methods; (ii) explicit threshold selection is no longer required; and (iii) different tail behaviours can be modelled within a unified parametric structure.

Let $X$ be a random variable following a two parameter GPD with scale parameter $\sigma>0$ and shape parameter $\xi \in \mathbb{R}$. The shape parameter $\xi$ is known as  the extreme value index (EVI) or the tail index and measures the heaviness of the right tail of the underlying distribution. The cumulative distribution function, denoted by $G$, is defined by

\[
G(x)=
\begin{cases}
1-\left(1+\xi x/\sigma\right)^{-1/\xi}, & \xi \neq 0,\\[6pt]
1-\exp(-x/\sigma), & \xi =0.
\end{cases}
\]

The support of the distribution depends on the value of $\xi$:
\[
x \in
\begin{cases}
[0,\infty), & \xi \ge 0,\\[6pt]
[0,-\sigma/\xi], & \xi <0.
\end{cases}
\]

The associated density function is
\[
g(x)=\frac{1}{\sigma}
\left(1+\xi x/\sigma\right)^{-1/\xi-1},
\]
defined for all $x$ satisfying $1+\xi x/\sigma>0$.

Although the GPD is defined for any $\xi\in\mathbb{R}$, the simulation study focuses on $\xi>0$, corresponding to heavy-tailed Pareto-type distributions. In the real-data applications, however, $\xi$ is estimated without imposing positivity.

The EGP distribution (EGPD) is constructed by composing a GPD with a distribution function supported on $[0,1]$. Let $U\sim\mathcal{U}[0,1]$ and let $G$ denote the GPD distribution function with scale and shape parameters $(\sigma,\xi)$. Given a distribution function $H$ on $[0,1]$, define a random variable
\[
X = G^{-1}\!\big(H^{-1}(U)\big).
\]
The cumulative distribution function of $X$ is
\[
F(x)=H\{G(x)\}.
\]
Provided that $H$ is differentiable, the density for the EGPD is given by:
\[
f(x)=h\{G(x)\}\,g(x), \qquad x\ge0,
\]
where $h$ and $g$ denote the densities of $H$ and $G$, respectively. Thus, the EGPD can be viewed as a deformation of the GPD, with the factor $h(G(x))$ modifying its behaviour while retaining the GPD as the underlying tail-generating distribution.

The role of $H$ can also be interpreted through inverse transform sampling. Whereas a standard GPD observation is obtained by applying $G^{-1}$ directly to a uniform random variable, the EGPD first transforms the uniform input through $H^{-1}$. The distribution function $H$ therefore controls the departure from the GPD, providing additional flexibility in the bulk of the distribution. To ensure that this additional flexibility does not alter the asymptotic right-tail behaviour, $H$ is required to satisfy the regularity condition introduced in the original EGPD formulation \citep{CoolArticle}:
\[
\lim_{v\to 0^+}\frac{1-H(1-v)}{v}=a,
\qquad a>0.
\]
This condition specifies the behaviour of $H$ near the upper endpoint of its support. In particular,
\[
1-H(1-v)\sim av
\qquad\text{as }v\to0^+,
\]
so that $H$ is asymptotically linear near $1$. Whenever $h$ is sufficiently regular, this is equivalently expressed as
\[
\lim_{u\to1^-}h(u)=a.
\]
Hence, near the upper boundary, $H$ behaves locally like a linear transformation, with its density converging to a positive constant.

This condition is what allows the EGPD to preserve the tail index of the underlying GPD. As $x$ becomes large, $G(x)$ approaches $1$, and the asymptotic linearity of $H$ implies that the transformation $H\{G(x)\}$ modifies the GPD tail only by a multiplicative factor. Consequently, the EGPD retains the same right-tail index as the GPD, while allowing departures from the GPD form in the body of the distribution.

The original formulation also imposes additional conditions on the behaviour of $H$ near the lower endpoint of $[0,1]$. Since these conditions are primarily related to the left-tail behaviour and are not required for the right-tail inference considered here, we do not impose them.

The authors in \cite{Extension} further extended the EGP class to a semi-parametric framework by estimating the perturbation distribution $H$ using Bernstein polynomials rather than pre-specifying its functional form. In this work, we extend this framework to accommodate right-censored data. In addition, we investigate an alternative semi-parametric specification by replacing the Bernstein polynomial representation with a more general sieve-based estimator \citep{Grenander1981}, aiming to increase flexibility in the estimation of the perturbation while maintaining theoretical tractability.

\section{Censored EGP models and estimation}

All models developed in this paper are formulated within the EGP framework introduced previously. Let $X$ denote the event time random variable and let $C$ denote the censoring time random variable. Under right censoring, the event time is not necessarily observed for every individual. Instead, the observed follow-up time is given by

$$
T=\min(X,C),
\qquad
\delta=\mathbf{1}(X\leq C),
$$

where $\delta\in{0,1}$ is the event indicator. Thus, $\delta=1$ indicates that the event occurs at time $T=X$ and is observed, whereas $\delta=0$ indicates that the event is right-censored at time $T=C$.

The statistical sample is given by
$
\{(T_i,\delta_i)\}_{i=1}^n,
$
where

$$
T_i=\min(X_i,C_i),
\qquad
\delta_i=\mathbf{1}(X_i\leq C_i),
\qquad i=1,\ldots,n,
$$

and $(T_i,\delta_i)$, $i=1,\ldots,n$, are independent and identically distributed realizations of the random pair $(T,\delta)$.

Within this framework, the EGP model is specified for the event time distribution, rather than for the censoring mechanism. Let \(Y\) denote the event time, with distribution function obtained by perturbing a generalized Pareto baseline. Specifically, let \(G_\theta\) denote a GPD with parameter vector \(\theta=(\sigma,\xi)\), and let \(H:[0,1]\to[0,1]\) be a distribution function acting as a transformation on the unit interval. The event time distribution is then defined by
\[
F_{\theta,H}(y)
=
H\!\big(G_\theta(y)\big),
\]
which can be interpreted as a deformation of the GPD induced by the perturbation \(H\). The censoring time is treated separately and is not assumed to follow the EGP model.

Three alternative specifications for the perturbation distribution $H$ are considered: a parametric Beta perturbation, a Bernstein polynomial estimator, and a histogram-based estimator. All three specifications are fitted using the same iterative estimation strategy introduced in \cite{Extension}, which is extended here to accommodate right-censored data. In particular, the censoring mechanism is incorporated into each step of the estimation procedure, while the EGP model remains specified for the event time distribution.

At each iteration, the current GPD estimate is used to construct the corresponding censored pseudo-sample, from which the perturbation distribution $H$ is estimated according to its particular specification. For Beta perturbation, the parameters are estimated by maximum likelihood, whereas Bernstein polynomial and histogram specifications are estimated using the procedures described in the corresponding subsections below. Thus, although the estimation of $H$ differs across specifications, all models are embedded within the same iterative scheme, allowing for a consistent comparison of the different perturbation structures under right censoring.

The estimation starts with a maximum-likelihood fit of a GPD to the censored sample, yielding an initial estimate $\widehat{\theta}^{(0)}=(\widehat{\sigma}^{(0)},\widehat{\xi}^{(0)})$. At iteration $k$, the event and censoring times are transformed to the unit interval according to
\[
Z_i^{(k)}
=
G_{\widehat{\theta}^{(k)}}(T_i),
\qquad i=1,\ldots,n.
\]
The resulting pseudo-sample retains the original censoring indicators, so that both the transformed event times and the transformed censoring times are used in the estimation of the perturbation. The distribution $H$ is then estimated from this censored pseudo-sample, yielding $\widehat{H}^{(k)}$. Importantly, the estimation of $H$ is not performed by maximum likelihood; rather, the appropriate procedure for the chosen perturbation specification is applied to the pseudo-sample, as described in the corresponding subsections below.

The fitted perturbation is subsequently used to transform the pseudo-sample back to the original scale:
\[
V_i^{(k)}
=
G_{\widehat{\theta}^{(k)}}^{-1}
\left\{
\widehat{H}^{(k)}
\left(
G_{\widehat{\theta}^{(k)}}(T_i)
\right)
\right\}.
\]
The updated pseudo-sample $\{(V_i^{(k)},\delta_i)\}_{i=1}^n$ is constructed simultaneously for both event and censored observations. Thus, when $\delta_i=1$, $V_i^{(k)}$ represents an updated event time, whereas when $\delta_i=0$, it represents the corresponding updated censoring time. A new censored-data GPD maximum-likelihood estimate is then obtained from this updated sample, producing $\widehat{\theta}^{(k+1)}$.

The iterations continue until the change in the estimated shape parameter satisfies
\[
\left|
\widehat{\xi}^{(k+1)}
-
\widehat{\xi}^{(k)}
\right|
<\varepsilon,
\qquad
\varepsilon=10^{-3}.
\]
The scale parameter is updated at every iteration but is not included in the convergence criterion. The complete iterative procedure is summarized in Algorithm~\ref{alg:censored-egpd}.

\begin{algorithm}
\caption{Iterative estimation of censored EGP models}
\label{alg:censored-egpd}
\begin{algorithmic}[1]
\Require $\{(T_i,\delta_i)\}_{i=1}^n$, $\mathcal{H}$, $\varepsilon=10^{-3}$
\Ensure $\widehat{\theta}=(\widehat{\sigma},\widehat{\xi})$, $\widehat{H}$

\State $\widehat{\theta}^{(0)}
\gets
\operatorname*{arg\,max}_{\theta}
\mathcal{L}_{\mathrm{GPD}}
\left(
\theta;\{(T_i,\delta_i)\}_{i=1}^n
\right)$
\State $k\gets0$

\Repeat

    \State
    \[
    Z_i^{(k)}
    \gets
    G_{\widehat{\theta}^{(k)}}(T_i),
    \qquad i=1,\ldots,n
    \]

    \State
    \[
    \mathcal{Z}^{(k)}
    \gets
    \left\{
    (Z_i^{(k)},\delta_i)
    \right\}_{i=1}^n
    \]

    \State
    \[
    \widehat{H}^{(k)}
    \gets
    \operatorname{Estimate}_{\mathcal{H}}
    \left(
    \mathcal{Z}^{(k)}
    \right)
    \]

    \State
    \[
    V_i^{(k)}
    \gets
    G_{\widehat{\theta}^{(k)}}^{-1}
    \left\{
    \widehat{H}^{(k)}
    \left(
    Z_i^{(k)}
    \right)
    \right\},
    \qquad i=1,\ldots,n
    \]

    \State
    \[
    \mathcal{V}^{(k)}
    \gets
    \left\{
    (V_i^{(k)},\delta_i)
    \right\}_{i=1}^n
    \]

    \State
    \[
    \widehat{\theta}^{(k+1)}
    \gets
    \operatorname*{arg\,max}_{\theta}
    \mathcal{L}_{\mathrm{GPD}}
    \left(
    \theta;\mathcal{V}^{(k)}
    \right)
    \]

    \State $k\gets k+1$

\Until{
$\left|
\widehat{\xi}^{(k)}
-
\widehat{\xi}^{(k-1)}
\right|
<\varepsilon$
}

\State \Return $\widehat{\theta}^{(k)},\widehat{H}^{(k-1)}$

\end{algorithmic}
\end{algorithm}

To the best of our knowledge, the behaviour of the Beta and Bernstein  EGP perturbations under right censoring has not previously been investigated, even under a fully likelihood-based estimation framework. Thus, in addition to proposing the histogram-based specification, this work extends the existing Beta and Bernstein EGP formulations to the right-censored setting.

\subsection{Beta perturbation}

The first specification corresponds to the parametric perturbation
\[
H^{Beta}(u)=u^{k},\qquad k>0,
\]
which induces a Beta$(k,1)$ distribution on \([0,1]\) and coincides with the original EGP formulation of \cite{CoolArticle}. The resulting model is
\[
F^{Beta}_{\theta,k}(t)
=\big(G_\theta(t)\big)^k,
\]
with density
\[
f^{Beta}_{\theta,k}(t)
=k\,\big(G_\theta(t)\big)^{k-1}g_\theta(t),
\]
and survival function
\[
S^{Beta}_{\theta,k}(t)
=1-\big(G_\theta(t)\big)^k.
\]

The regularity condition of the EGP definition requires asymptotic linearity of the perturbation near the upper endpoint. Since
\[
H^{Beta}(1-v)=(1-v)^k=1-kv+o(v),
\]
we obtain
\[
\frac{1-H^{Beta}(1-v)}{v}\to k>0,
\]
so the perturbation preserves the extreme-value index and the tail behaviour remains governed entirely by the GPD parameter \(\xi\).

The parameter \(k\) is estimated within the iterative framework described above using likelihood maximization under right censoring via differential evolution \citep{storn1997differential}. Throughout the remainder of the paper, this specification will be referred to interchangeably as the Beta model, the parametric model, or simply the EGP model.

\subsection{Bernstein polynomial perturbation}\label{sub:Bernstein}

The Bernstein polynomial may be considered to yield a smooth nonparametric estimator for the perturbation distribution $H(t)$, while preserving the distributional constants required by the EGPD construction. This approach, considered by \cite{Extension}, is adopted for a more general setting considering right-censored data.

For a general perturbation $H(t)$, consider the Bernstein approximation of
degree $ m > 1$ given by

\[
H_m^{Bern}(t)
=\sum_{k=0}^{m}
H\!\left(\frac{k}{m}\right)
\binom{m}{k}t^k(1-t)^{m-k}.
\]
After differentiation, we obtain
\[
h_m(t)
=\sum_{k=1}^{m}\omega_{k,m}\,
\beta_{k,m-k+1}(t),
\]
where $\beta_{a,b}(t)$ is the probability density function of a $Beta(a, b)$ distribution.
Therefore, $h_m(t)$ may be seen as a mixture of Beta densities, a smooth approximation of the perturbation density $H'(t)$.

In \cite{Extension}, the weights $\omega_{k,m}$ are estimated by the jumps of the empirical distribution function. To accommodate the censored observations, we replace it by jumps of the Kaplan–Meier estimator, given by
\[
\hat\omega_{k,m}
=
\hat S^{(m)}\!\left(\frac{k-1}{m}\right)
-
\hat S^{(m)}\!\left(\frac{k}{m}\right),
\qquad
k=1,\ldots,m.
\]

with $\hat S^{(m)}(\cdot)$ computed based on

\[
Z_i=G_\theta(T_i),
\qquad i=1,\ldots,n,
\]
the current GPD with parameter $\theta$.

Heavy censoring may produce a flat Kaplan--Meier curve near the boundary, leading to a vanishing final weight. Since preservation of the EGP tail behaviour requires \(\omega_{m,m}>0\), the normalization strategy proposed in \cite{Extension} is adopted. When the last increment is zero, it is replaced by
\[
\tilde\omega_{m,m}
=1-\widehat H_m^{Bern}\!\left(1-\frac{1}{m}\right),
\]
followed by renormalization,
\[
\omega_{k,m}
=\frac{\tilde\omega_{k,m}}
{\sum_{j=1}^{m}\tilde\omega_{j,m}},
\qquad k=1,\ldots,m,
\]
ensuring
\[
\omega_{k,m}\ge0,\qquad
\sum_{k=1}^{m}\omega_{k,m}=1,\qquad
\omega_{m,m}>0.
\]
Consequently, the perturbation remains flexible in the body of the distribution while the right-tail index continues to be determined by the GPD component.

\subsection{Histogram-based perturbation}

The histogram sieve specification, following the sieve approach to nonparametric density estimation introduced by \citep{Grenander1981}, provides a fully nonparametric alternative based on piecewise-constant approximations to the density. In contrast to the Bernstein approach, no smoothing is introduced; the perturbation distribution is constructed directly from probability masses derived from the Kaplan--Meier survival function.

Consider the regular partition \(j\) for grid points,  \(k\) for bins
\[
0=t_0<t_1<\cdots<t_m=1,
\qquad t_j=\frac{j}{m},\,\, j=0, \ldots, m.
\]
With \(\hat S^{(m)}\) the Kaplan--Meier estimator of the survival function, introduced in section \ref{sub:Bernstein}, the
raw histogram masses are initially defined as the estimated probability mass contained in each bin
\[
\tilde\omega_{k,m}
=\hat S^{(m)}(t_{k-1})-\hat S^{(m)}(t_k),
\qquad k=1,\ldots,m.
\]

Finite samples and censoring may lead to empty upper bins and, in particular, to a vanishing last increment, which is problematic for ensuring a proper density representation. To address this, we identify the last nonempty bin and merge the remaining tail mass. Let
\[
r=\max\{k\in\{1,\ldots,m\}:\tilde\omega_{k,m}>0\}
\]
be the index of the last bin containing positive mass. All bins above \(r\) are then merged, producing the aggregated tail mass
\[
M_r=\sum_{k=r}^{m}\tilde\omega_{k,m}.
\]

An adaptive partition 
\(
0=s_0<s_1<\cdots<s_r=1
\)
is constructed by retaining the original grid up to \(t_{r-1}\) and enlarging the final interval to cover the remaining support,
\[
s_j=t_j,\qquad j=0,\ldots,r-1,
\qquad\text{and}\qquad s_r=1.
\]
For \(k>r\), \(s_k\) is left undefined.

Denoting the bin widths by \(\Delta_k=s_k-s_{k-1}\), the histogram masses after merging are defined as
\[
\omega_k=
\begin{cases}
\tilde\omega_{k,m}, & k<r,\\
\hat S^{(m)}(t_r)=\hat S^{(m)}(t_m)
= M_r + \Big(1 - \sum_{j=1}^{r-1}\tilde\omega_{j,m}\Big), & k=r,
\end{cases}
\]
where the equality \(\widehat S^{KM}(t_r)=\widehat S^{KM}(t_m)\) follows from the fact that no Kaplan--Meier jumps occur in the interval \((t_r,t_m]\), so that the survival function remains constant over the merged tail region. The additional term \(1 - \sum_{j=1}^{r-1}\tilde\omega_{k,m}\)  enforces global mass conservation, ensuring that the histogram representation remains a proper probability distribution.

This construction directly replaces the previously aggregated tail mass obtained after merging, following a similar strategy to that applied in \cite{Extension}.

The histogram density is obtained via the standard mass--over--width construction,
\[
\widehat h^{Hist}(t)
=\sum_{k=1}^{r}
\frac{\omega_k}{\Delta_k}
\mathbf{1}_{(s_{k-1},s_k]}(t),
\]
which defines a proper density on \([0,1]\) while preserving the total Kaplan--Meier probability mass.

In a neighborhood of the boundary, this construction implies
\[
1-\widehat H^{Hist}(t)
=\frac{\omega_r}{\Delta_r}(1-t),
\qquad t\uparrow1,
\]
and therefore
\[
\lim_{t\uparrow1}
\frac{1-\widehat H^{Hist}(t)}{1-t}
=\frac{\omega_r}{\Delta_r}>0.
\]
Hence the regularity condition is automatically satisfied and the EGP tail index remains unchanged.

In all cases, the perturbation affects only the body of the distribution, while the extreme right tail is entirely governed by the GPD component.

Regarding the three methods above described, they differ in their degree of parametrization, smoothness, and flexibility, while sharing the same EGP tail structure. The Beta specification corresponds to the original parametric EGP formulation introduced in \cite{CoolArticle}. The Bernstein specification provides a smooth semi-parametric extension of the EGP framework developed in \cite{Extension}. Finally, the histogram sieve considered in this work provides a flexible semi-parametric representation of the perturbation and constitutes the main methodological contribution of this paper. A summary of the main characteristics of the three specifications is provided in Table~\ref{tab:perturbation_comparison}.

\begin{table}[htbp]
\centering
\caption{Comparison of the perturbation specifications.
P: parametric; SP: semi-parametric; S: smooth; PC: piecewise constant.
The last column indicates the source of the formulation and the contribution of the present work}
\label{tab:perturbation_comparison}
\begin{tabular}{lccc}
\hline
\textbf{Specification} & \textbf{Type} & \textbf{Smoothness} & \textbf{Origin} \\
\hline
Beta       & P  & S  & \cite{CoolArticle}; censored extension  \\
Bernstein  & SP & S  & \cite{Extension}; censored extension \\
Histogram  & SP & PC & Proposed here \\
\hline
\end{tabular}
\end{table}

\section{Simulation study}

A Monte Carlo simulation study was conducted to evaluate the finite-sample performance of the three EGP specifications considered in this work: the parametric Beta perturbation, the Bernstein polynomial estimator, and the proposed histogram-based estimator. The objective is to assess the estimator performance under different tail regimes, censoring mechanisms, sample sizes, and levels of nonparametric flexibility.

\subsection{Simulation design and performance measures}

For each replication, event times were independently generated from a GPD,

$$
X \sim \mathrm{GPD}(\sigma,\xi),
$$

with location parameter fixed at $\mu=0$, scale parameter $\sigma=1$, and shape parameter $\xi$. Independently, right-censoring times were generated from a second GPD,

$$
C \sim \mathrm{GPD}(\sigma_c,\xi_c),
$$

where the censoring scale parameter $\sigma_c$ and tail index $\xi_c$ were varied across simulation scenarios. The observed data were then constructed as $(T_i,\delta_i)$, where $T_i=\min(X_i,C_i)$ and $\delta_i=\mathbf{1}(X_i\leq C_i)$, following the notation introduced before.

The simulation design considered four different values for the event tail index,
\[
\xi\in\{0.5,1.0,1.5,2.0\},
\]
and three censoring tail regimes,
\[
\xi_c\in\{-1,0,1\}.
\]
Sample sizes were taken as
\[
n\in\{50,100,500,1000\}.
\]

For the nonparametric perturbation estimators, the number of basis components or bins was also varied according to
\[
m\in\{10,20,80,140\}.
\]
The parametric Beta perturbation does not depend on $m$, but the same simulation scenarios were retained for comparability across methods.

The censoring scale parameter $\sigma_c$ was selected to produce approximately $25\%$, $50\%$, and $75\%$ censoring proportions, depending on the combination of $(\xi,\xi_c)$. The values used in the simulation are summarized in Table~\ref{tab:censoring_scales}.

\begin{table}[H]
\centering
\caption{Censoring scale parameters $\sigma_c$ used in the simulation
study for each combination of event tail index $\xi$ and censoring tail index $\xi_c$. The three values correspond approximately to
light, moderate, and heavy censoring scenarios}
\label{tab:censoring_scales}
\begin{tabular}{cccc}
\toprule
$\xi$ & $\xi_c=-1$ & $\xi_c=0$ & $\xi_c=1$ \\
\midrule
0.5 & $(0.7,\,2,\,6)$ & $(0.4,\,1.3,\,4.5)$ & $(0.2,\,0.8,\,7.5)$ \\
1.0 & $(0.7,\,2.5,\,9.5)$ & $(0.4,\,1.7,\,7)$ & $(0.2,\,1,\,5)$ \\
1.5 & $(0.8,\,3,\,15)$ & $(0.5,\,2,\,11)$ & $(0.2,\,1.3,\,8)$ \\
2.0 & $(0.9,\,4,\,24)$ & $(0.5,\,2.5,\,18)$ & $(0.3,\,1.6,\,12.5)$ \\
\bottomrule
\end{tabular}
\end{table}

For each parameter configuration, $1,000$ independent Monte Carlo replications were performed. In every replication, the three EGP specifications were fitted to the censored sample using the iterative estimation procedure described previously.

The simulation analysis focuses on both the censoring mechanism and the quality of parameter estimation. Since the number of bins $m$ affects only the perturbation estimation step and not the data-generating process, the censoring proportion is identical across all values of $m$ within a given scenario. For each configuration, the empirical mean of the censoring proportion, the median estimated scale parameter $\widehat{\sigma}$, the median estimated tail index $\widehat{\xi}$, and the mean absolute errors (MAEs) and bias of $\widehat{\sigma}$ and $\widehat{\xi}$ were recorded.

For a parameter $\theta = (\sigma,\xi)$, the MAE for a given simulation configuration is defined as
\[
\operatorname{MAE}(\widehat{\theta})
=
\frac{1}{R}
\sum_{b=1}^{R}
\left|
\widehat{\theta}^{(b)}-\theta
\right|,
\]
where $R=1,000$ denotes the number of Monte Carlo replications, $\widehat{\theta}^{(b)}$ is the estimate obtained in replication $b$, and $\theta$ is the corresponding true parameter value.

To compare the three EGP specifications across the simulation configurations, we consider both the \emph{overall proportion of lowest MAE} and the \emph{overall proportion of lowest absolute bias}. Let $\mathcal{K}$ denote the set of simulation configurations being compared and let $M_{jk}$ denote the MAE of model $j$ in configuration $k$. The overall proportion of lowest MAE for model $j$ is defined as
\[
P^{\mathrm{MAE}}_j
=
\frac{1}{|\mathcal{K}|}
\sum_{k\in\mathcal{K}}
\mathbf{1}
\left\{
M_{jk}
=
\min_{\ell\in\mathcal{M}} M_{\ell k}
\right\},\qquad j\in \mathcal{M},
\]
where
\[
\mathcal{M}
=
\{\text{Beta},\text{Bernstein},\text{Histogram}\}.
\]
Thus, $P^{\mathrm{MAE}}_j$ represents the proportion of simulation configurations in which model $j$ achieves the lowest MAE among the three competing specifications. Larger values of $P^{\mathrm{MAE}}_j$ indicate that a model performs best more frequently across the considered simulation scenarios.

For bias, let $\widehat{B}_{jk}$ denote the empirical bias of model $j$ in configuration $k$, calculated as the average signed estimation error across the Monte Carlo replications within that configuration. For a parameter $\theta$, this is given by
\[
\widehat{B}_{jk}
=
\frac{1}{R_k}
\sum_{r=1}^{R_k}
\left(
\widehat{\theta}_{jk}^{(r)}-\theta_k
\right),
\]
where $R_k$ is the number of replications for configuration $k$, $\widehat{\theta}_{jk}^{(r)}$ is the estimate obtained by model $j$ in replication $r$, and $\theta_k$ is the true parameter value. To identify the model with the smallest bias, we compare the absolute values of these empirical biases, since a bias closer to zero indicates less systematic over- or underestimation. The overall proportion of lowest absolute bias for model $j$ is therefore defined as
\[
P^{\mathrm{Bias}}_j
=
\frac{1}{|\mathcal{K}|}
\sum_{k\in\mathcal{K}}
\mathbf{1}
\left\{
|\widehat{B}_{jk}|
=
\min_{\ell\in\mathcal{M}}
|\widehat{B}_{\ell k}|
\right\},\qquad j\in \mathcal{M}.
\]
Thus, $P^{\mathrm{Bias}}_j$ represents the proportion of simulation configurations in which model $j$ has the bias closest to zero among the three competing specifications. Larger values of $P^{\mathrm{Bias}}_j$ indicate that a model exhibits the smallest systematic deviation from the true parameter value more frequently across the considered simulation scenarios.

\subsection{Simulation results}

Tables~\ref{tab:overall_winner_mae} -- \ref{tab:summary 03 for censoring}
summarize the comparative performance of the three models in terms of MAE and the proportion of configurations in which each model attains the lowest MAE. Complementing these results, Tables~\ref{tab:tab:bias_winner_overall} -- \ref{tab:tab:bias_winner_censoring}
report the corresponding proportions for the model with the lowest absolute bias. Given their size and level of detail, the full simulation results are reported in Tables~\ref{tab:tab:sim_beta} -- \ref{tab:tab:sim_hist}
in the Supplementary Material.

Table~\ref{tab:overall_winner_mae} shows that the Beta perturbation performs best in estimating the tail index $\xi$, followed by the histogram and then the Bernstein specification. For the scale parameter $\sigma$, however, the ranking is reversed: the histogram performs best, followed by the Bernstein model, while the Beta specification performs worst.

\begin{table}[ht]
\centering
\caption{Overall proportion of lowest MAE for $\xi$ and scale across all configurations}
\label{tab:overall_winner_mae}
\begin{tabular}{lrr}
\toprule
Model & Prop. ($\xi$) & Prop. (scale) \\
\midrule
Beta       & 0.4094 & 0.1906 \\
Bernstein  & 0.2557 & 0.3818 \\
Histogram  & 0.3349 & 0.4276 \\
\bottomrule
\end{tabular}
\end{table}

When conditioning on tail heaviness (Table~\ref{tab:summary 01 for tail}), this ordering remains broadly stable. Heavier tails tend to improve the relative performance of the Beta model for $\xi$, while further weakening its performance for $\sigma$. The effect of sample size (Table~\ref{tab:summary 02 for sample}) is more pronounced. For small samples, the Bernstein model performs slightly better overall for both parameters, with the histogram close behind and the Beta model clearly worse. As the sample size increases, the pattern observed in the full sample re-emerges: the Beta model improves significantly for $\xi$, while the histogram becomes increasingly competitive for estimating $\sigma$.

\begin{table}[ht]
\centering
\caption{Proportion of lowest MAE for $\xi$ and scale across tail indices}
\label{tab:summary 01 for tail}
\begin{tabular}{rlrr}
\toprule
$\xi$ & Model & Prop. ($\xi$) & Prop. (scale) \\
\midrule
0.5 & Beta      & 0.368 & 0.214 \\
0.5 & Bernstein & 0.317 & 0.381 \\
0.5 & Histogram & 0.316 & 0.405 \\
\addlinespace
1.0 & Beta      & 0.400 & 0.197 \\
1.0 & Bernstein & 0.268 & 0.377 \\
1.0 & Histogram & 0.332 & 0.426 \\
\addlinespace
1.5 & Beta      & 0.429 & 0.182 \\
1.5 & Bernstein & 0.229 & 0.383 \\
1.5 & Histogram & 0.341 & 0.435 \\
\addlinespace
2.0 & Beta      & 0.440 & 0.168 \\
2.0 & Bernstein & 0.209 & 0.387 \\
2.0 & Histogram & 0.351 & 0.445 \\
\bottomrule
\end{tabular}
\end{table}

\begin{table}[ht]
\centering
\caption{Proportion of lowest MAE for $\xi$ and scale across sample sizes}
\label{tab:summary 02 for sample}
\begin{tabular}{rlrr}
\toprule
$n$ & Model & Prop. ($\xi$) & Prop. (scale) \\
\midrule
50   & Beta      & 0.285 & 0.177 \\
50   & Bernstein & 0.371 & 0.429 \\
50   & Histogram & 0.344 & 0.394 \\
\addlinespace
100  & Beta      & 0.358 & 0.165 \\
100  & Bernstein & 0.306 & 0.432 \\
100  & Histogram & 0.335 & 0.403 \\
\addlinespace
500  & Beta      & 0.454 & 0.191 \\
500  & Bernstein & 0.221 & 0.373 \\
500  & Histogram & 0.325 & 0.436 \\
\addlinespace
1000 & Beta      & 0.465 & 0.216 \\
1000 & Bernstein & 0.195 & 0.329 \\
1000 & Histogram & 0.340 & 0.455 \\
\bottomrule
\end{tabular}
\end{table}

Finally, Table~\ref{tab:summary 03 for censoring} highlights the effect of censoring. Low censoring is particularly favorable for the histogram estimator, while it is less favorable for the Bernstein specification. As censoring increases, the performance of the histogram deteriorates, while the Bernstein estimator improves relative to the other specifications. The Beta model remains relatively stable in estimating $\xi$ across censoring regimes but is consistently weaker for scale estimation.

\begin{table}[ht]
\centering
\caption{Proportion of lowest MAE for $\xi$ and scale across censoring levels}
\label{tab:summary 03 for censoring}
\begin{tabular}{llrr}
\toprule
Censoring & Model & Prop. ($\xi$) & Prop. (scale) \\
\midrule
Low    & Beta      & 0.421 & 0.185 \\
Low    & Bernstein & 0.159 & 0.305 \\
Low    & Histogram & 0.421 & 0.510 \\
\addlinespace
Medium & Beta      & 0.408 & 0.194 \\
Medium & Bernstein & 0.278 & 0.400 \\
Medium & Histogram & 0.314 & 0.406 \\
\addlinespace
High   & Beta      & 0.399 & 0.192 \\
High   & Bernstein & 0.331 & 0.441 \\
High   & Histogram & 0.270 & 0.367 \\
\bottomrule
\end{tabular}
\end{table}

The bias results provide a complementary perspective on these findings. Table~\ref{tab:tab:bias_winner_overall} shows that the histogram specification has the highest proportion of configurations with the smallest absolute bias for both parameters, with proportions of $0.647$ for $\xi$ and $0.590$ for $\sigma$. The Bernstein model follows, with proportions of $0.263$ and $0.340$, respectively, whereas the Beta model has the lowest proportions, $0.090$ for $\xi$ and $0.071$ for $\sigma$. Thus, although the MAE results highlight particular advantages of the Beta specification for tail-index estimation, the bias results indicate a broader advantage of the histogram specification in terms of systematic deviation from the true parameter values.

\begin{table}

\caption{\label{tab:tab:bias_winner_overall}Overall proportion of lowest bias for $\xi$ and scale across all configurations}
\centering
\begin{tabular}[t]{lrr}
\toprule
Model & Proportion ($\xi$) & Proportion ($\sigma$)\\
\midrule
Beta & 0.0897 & 0.0705\\
Bernstein & 0.2628 & 0.3397\\
Histogram & 0.6474 & 0.5897\\
\bottomrule
\end{tabular}
\end{table}

This advantage of the histogram specification is also evident across tail indices (Table~\ref{tab:tab:bias_winner_tail}). The histogram model is the most frequent winner for both parameters at all tail indices. For $\xi$, its proportion increases from $0.564$ at $\xi=0.5$ to $0.692$ at $\xi=1.5$ and $\xi=2.0$, while the Beta model remains comparatively weak, with proportions between $0.077$ and $0.128$. For $\sigma$, the histogram model dominates at $\xi=0.5$ and $\xi=1.0$, whereas the Bernstein model becomes more competitive under heavier tails, attaining a proportion of $0.385$ at both $\xi=1.5$ and $\xi=2.0$. The Beta model is rarely the specification with the smallest absolute bias for $\sigma$ in these settings.

\begin{table}

\caption{\label{tab:tab:bias_winner_tail}Proportion of lowest bias for $\xi$ and scale across tail indices}
\centering
\begin{tabular}[t]{rlrr}
\toprule
$\xi$ & Model & Proportion ($\xi$) & Proportion ($\sigma$)\\
\midrule
0.5 & Beta & 0.128 & 0.154\\
0.5 & Bernstein & 0.308 & 0.308\\
0.5 & Histogram & 0.564 & 0.538\\
\addlinespace
1.0 & Beta & 0.077 & 0.128\\
1.0 & Bernstein & 0.282 & 0.282\\
1.0 & Histogram & 0.641 & 0.590\\
\addlinespace
1.5 & Beta & 0.077 & 0.000\\
1.5 & Bernstein & 0.231 & 0.385\\
1.5 & Histogram & 0.692 & 0.615\\
\addlinespace
2.0 & Beta & 0.077 & 0.000\\
2.0 & Bernstein & 0.231 & 0.385\\
2.0 & Histogram & 0.692 & 0.615\\
\bottomrule
\end{tabular}
\end{table}

The dependence on sample size is more nuanced (Table~\ref{tab:tab:bias_winner_sample}). For the smallest sample size, $n=50$, the Bernstein model has the largest proportion of configurations with the smallest absolute bias for both $\xi$ and $\sigma$, with proportions of $0.583$ and $0.667$, respectively, while the remaining configurations are won by the histogram model. At $n=100$, the histogram becomes dominant for $\xi$, with a proportion of $0.722$, whereas the Bernstein model remains more competitive for $\sigma$, attaining a proportion of $0.472$. For $n=500$ and $n=1000$, the histogram model is the most frequent winner for both parameters. At $n=1000$, in particular, it achieves proportions of $0.688$ for $\xi$ and $0.792$ for $\sigma$. These results suggest that the histogram specification becomes increasingly competitive in terms of bias as the sample size grows, whereas the Bernstein specification has a notable advantage in the smallest samples.

\begin{table}

\caption{\label{tab:tab:bias_winner_sample}Proportion of lowest bias for $\xi$ and scale across sample sizes}
\centering
\begin{tabular}[t]{rlrr}
\toprule
$n$ & Model & Proportion ($\xi$) & Proportion ($\sigma$)\\
\midrule
50 & Beta & 0.000 & 0.000\\
50 & Bernstein & 0.583 & 0.667\\
50 & Histogram & 0.417 & 0.333\\
\addlinespace
100 & Beta & 0.000 & 0.111\\
100 & Bernstein & 0.278 & 0.472\\
100 & Histogram & 0.722 & 0.417\\
\addlinespace
500 & Beta & 0.083 & 0.125\\
500 & Bernstein & 0.250 & 0.229\\
500 & Histogram & 0.667 & 0.646\\
\addlinespace
1000 & Beta & 0.208 & 0.021\\
1000 & Bernstein & 0.104 & 0.188\\
1000 & Histogram & 0.688 & 0.792\\
\bottomrule
\end{tabular}
\end{table}

Censoring produces an especially pronounced change in the relative bias performance (Table~\ref{tab:tab:bias_winner_censoring}). Under low censoring, the histogram specification overwhelmingly has the smallest absolute bias for both parameters, with proportions of $0.923$ for both $\xi$ and $\sigma$. Under medium censoring, the histogram remains the most frequent winner, with proportions of $0.615$ for $\xi$ and $0.654$ for $\sigma$. Under high censoring, however, the Bernstein model becomes substantially more competitive and is the most frequent winner for both parameters, with proportions of $0.596$ for $\xi$ and $0.654$ for $\sigma$. The corresponding histogram proportions fall to $0.404$ for $\xi$ and $0.192$ for $\sigma$. The Beta specification remains comparatively weak across the censoring regimes, particularly for $\xi$ under high censoring, where it has no configurations with the smallest absolute bias.

\begin{table}

\caption{\label{tab:tab:bias_winner_censoring}Proportion of lowest bias for $\xi$ and scale across censoring levels}
\centering
\begin{tabular}[t]{llrr}
\toprule
Censoring & Model & Proportion ($\xi$) & Proportion ($\sigma$)\\
\midrule
Low & Beta & 0.058 & 0.000\\
Low & Bernstein & 0.019 & 0.077\\
Low & Histogram & 0.923 & 0.923\\
\addlinespace
Medium & Beta & 0.212 & 0.058\\
Medium & Bernstein & 0.173 & 0.288\\
Medium & Histogram & 0.615 & 0.654\\
\addlinespace
High & Beta & 0.000 & 0.154\\
High & Bernstein & 0.596 & 0.654\\
High & Histogram & 0.404 & 0.192\\
\bottomrule
\end{tabular}
\end{table}

In addition to these comparisons, Tables~\ref{tab:tab:sim_bernstein} and \ref{tab:tab:sim_hist} show that larger values of $m$ are generally associated with higher MAE, although this effect becomes less pronounced at larger sample sizes. This suggests that $m$ should be chosen as small as possible while remaining sufficiently large for the perturbation model to be identifiable, particularly when the sample size is limited.

Overall, the MAE and bias results provide complementary evidence on the relative performance of the three specifications. The MAE results reveal clear parameter- and scenario-dependent differences, with the Beta specification particularly effective for $\xi$ and the histogram specification generally more competitive for $\sigma$. The bias results, in contrast, reveal a more pronounced and consistent advantage for the histogram specification across a broad range of configurations. Nevertheless, the Bernstein model can become preferable under specific challenging conditions, particularly small samples and high censoring. These findings suggest that the relative performance of the three specifications depends not only on the parameter of interest but also on the characteristics of the simulation scenario.

\subsection{Main conclusions}

Overall, the simulation results indicate that no single EGP specification dominates across all parameter-estimation tasks and simulation scenarios. Rather, the MAE and bias results highlight different aspects of the estimators' performance. In terms of MAE, the Beta specification is particularly effective for estimating the tail index $\xi$, especially as the sample size increases and under heavier tails, whereas the histogram specification is generally more competitive for scale estimation. When systematic deviation from the true values is considered, however, the histogram specification exhibits the strongest overall performance, achieving the highest proportion of configurations with the smallest absolute bias for both $\xi$ and $\sigma$.

The relative performance also depends strongly on the simulation setting. The histogram specification performs particularly well across different tail indices and becomes increasingly competitive as the sample size grows. The Bernstein specification, although less dominant overall, can provide substantial advantages in small samples and under high censoring. In particular, it becomes the most frequent winner in terms of absolute bias for both parameters under high censoring. The Beta specification, despite its strong MAE performance for $\xi$, is less frequently the model with the smallest absolute bias.

Taken together, these results highlight the complementary strengths of the three EGP specifications. The choice of perturbation model should therefore be guided by the inferential objective and by the characteristics of the data, including sample size, tail heaviness, and censoring intensity. The results also indicate that greater flexibility does not necessarily translate into better estimation: increasing $m$ can increase MAE, particularly when the sample size is limited. Thus, an appropriate balance between flexibility and stability is important for reliable EGPD estimation.

\section{Real data application}

\subsection{Bladder cancer recurrence data}

As an illustrative application, we consider the bladder cancer dataset available in the \texttt{survival} package in \textsf{R}, which originates from a randomized clinical trial on Stage~I tumors \citep{Byar1980}. We focus on the placebo and thiotepa groups and model the time to first recurrence. The sample sizes for the control and treatment groups are 47 and 38, respectively, with corresponding censoring proportions of approximately 38\% and 53\%.

Four models were fitted: the three censored EGP specifications proposed in this work and an exponential cure rate model included for comparison purposes. For the semi-parametric EGP models, the complexity parameter is fixed at $10$, corresponding to the number of bins in the histogram representation and, equivalently, to the degree of the Bernstein polynomial approximation. This value was selected based on the simulation study. Model performance is assessed through comparison with the Kaplan–Meier estimator and by examining the stability of parameter estimates across model specifications.

\begin{figure}[H]
    \centering
    \begin{subfigure}[t]{0.48\linewidth}
        \centering
        \includegraphics[width=\linewidth]{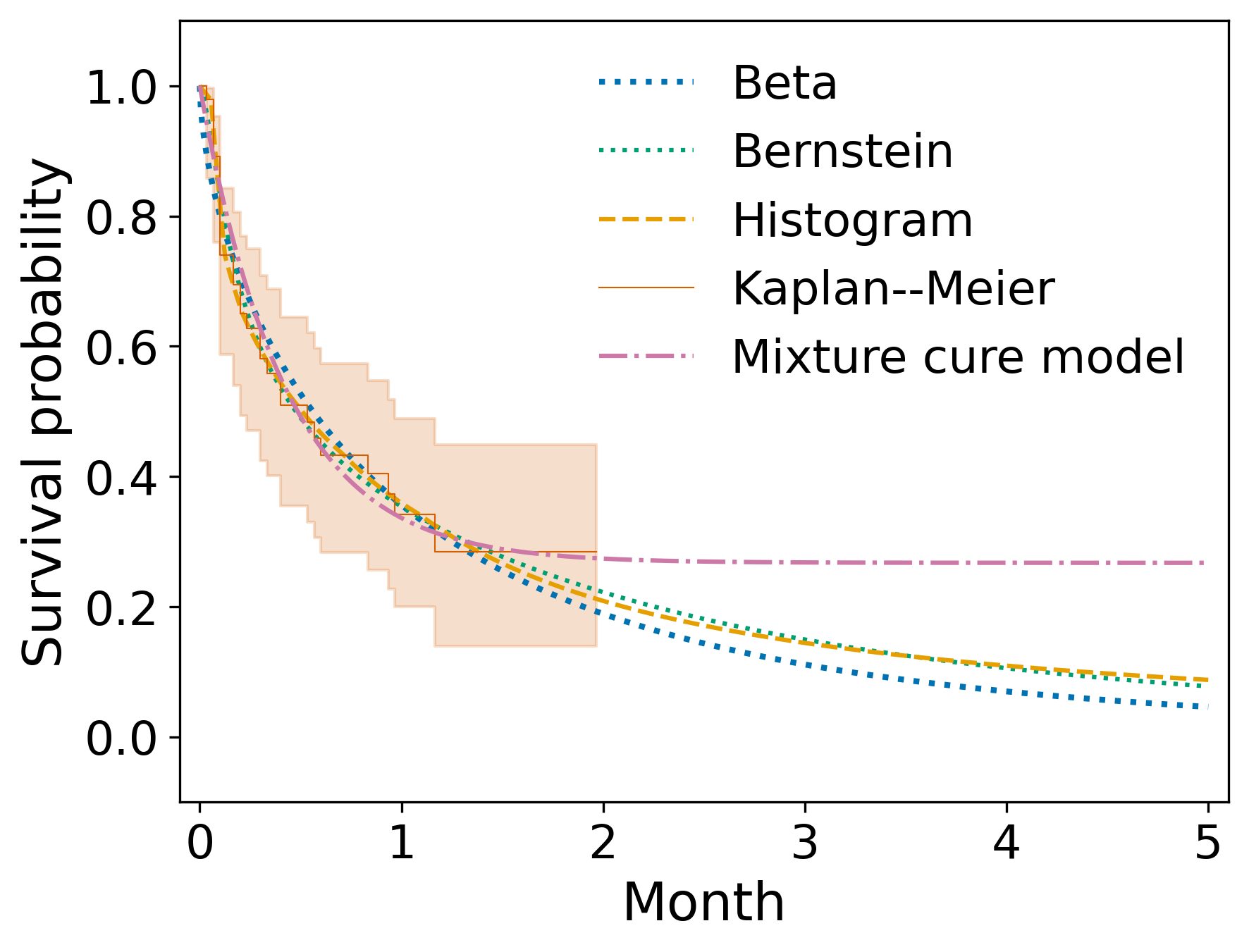}
        \caption{Control}
    \end{subfigure}
    \hfill
    \begin{subfigure}[t]{0.48\linewidth}
        \centering
        \includegraphics[width=\linewidth]{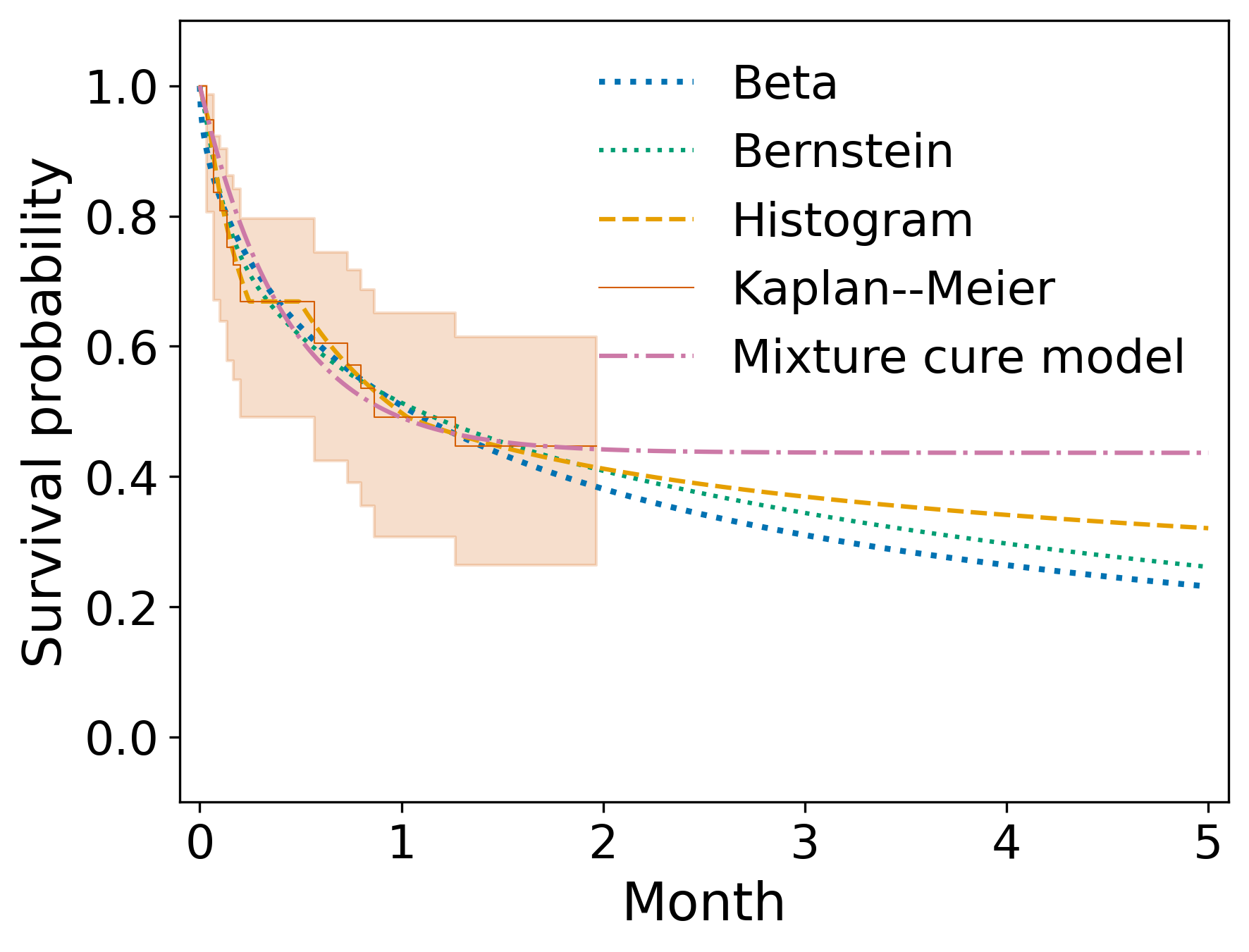}
        \caption{Treatment}
    \end{subfigure}
    \caption{Estimated survival functions for the bladder dataset}
    \label{fig:bladder}
\end{figure}

Figure~\ref{fig:bladder} shows the Kaplan--Meier estimator with confidence bands together with the fitted models. All specifications agree closely within the observed follow-up period, remaining inside the Kaplan--Meier bands. The cure model lies systematically above the others due to its long-term plateau assumption.

\begin{table}[htbp]
\centering
\caption{Estimated EVI $\xi$ and scale parameter for the control and treatment groups}
\label{tab:xi_scale}
\begin{tabular}{llcc}
\hline
Group & EGP  specification & $\hat{\xi}$ & $\hat{\sigma}$ \\
\hline
Control   & Beta       & 0.2305 & 1.4169 \\
          & Bernstein  & 0.5045 & 0.5226 \\
          & Histogram  & 0.8907 & 0.5044 \\
\hline
Treatment & Beta       & 1.3171 & 2.7149 \\
          & Bernstein  & 1.0463 & 0.4129 \\
          & Histogram  & 3.5390 & 0.3430 \\
\hline
\end{tabular}
\end{table}

Although the EGP specifications are nearly indistinguishable in the data-supported region, they diverge substantially in the tail. This difference is driven by the estimated EVI $\xi$. As reported in Table~\ref{tab:xi_scale}, for the control group, the Beta, Bernstein, and histogram models yield estimates of approximately $0.23$, $0.50$, and $0.89$, respectively. A similar pattern is observed for the treatment group, with estimates of $1.32$, $1.05$, and $3.54$, respectively.

\begin{figure}[H]
    \centering
    \begin{subfigure}[t]{0.48\linewidth}
        \centering
        \includegraphics[width=\linewidth]{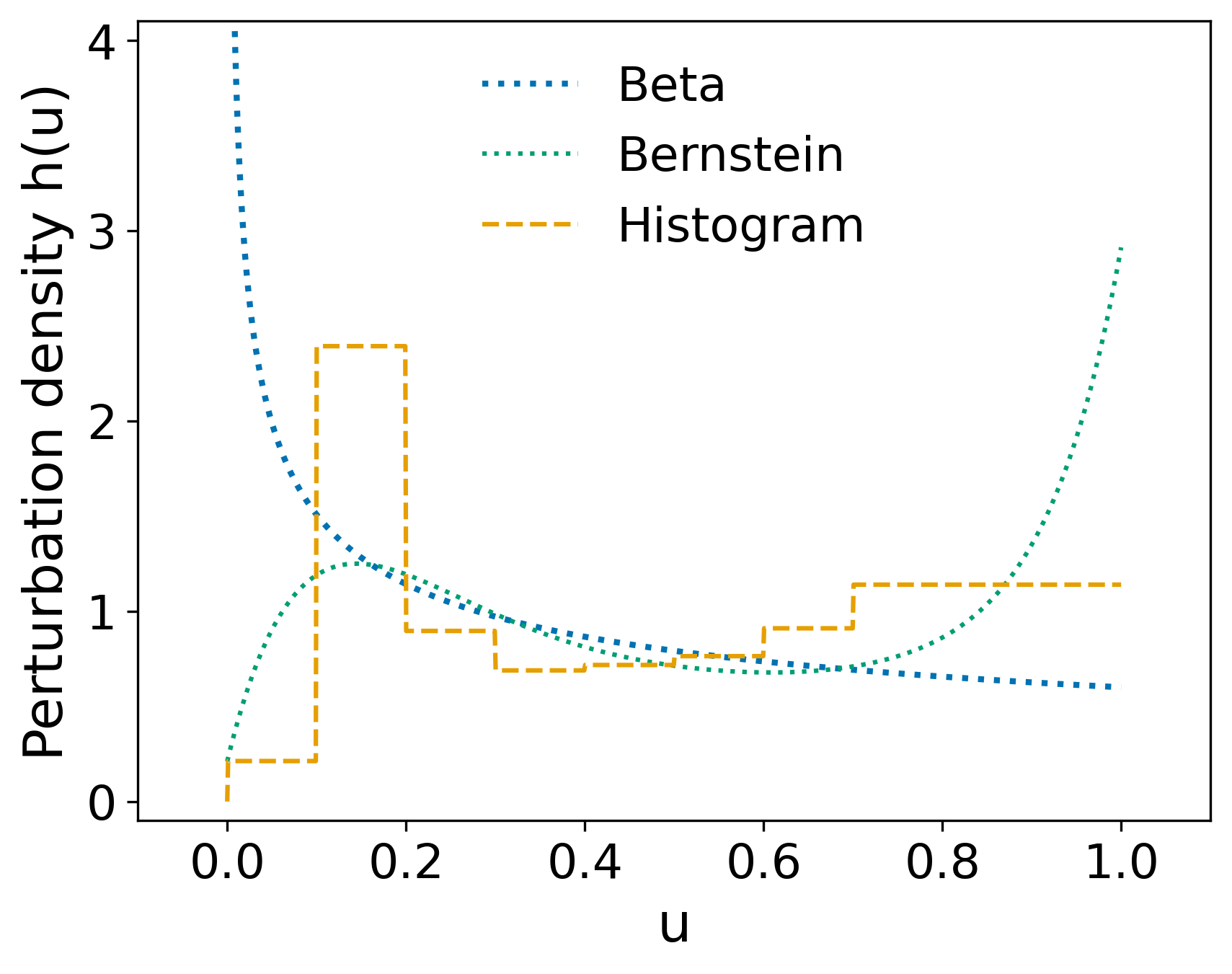}
        \caption{Control}
    \end{subfigure}
    \hfill
    \begin{subfigure}[t]{0.48\linewidth}
        \centering
        \includegraphics[width=\linewidth]{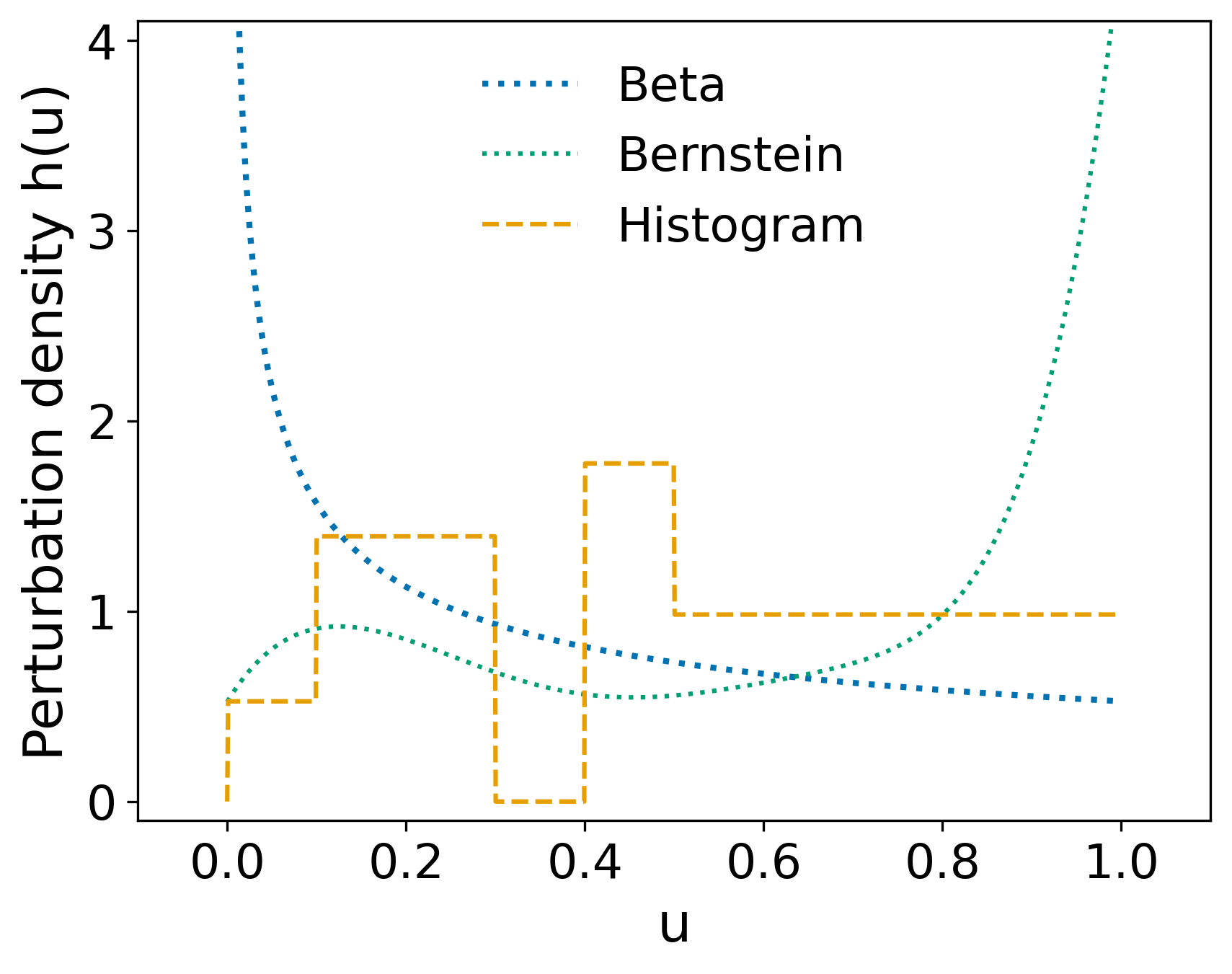}
        \caption{Treatment}
    \end{subfigure}
    \caption{Estimated perturbation densities for the bladder dataset}
    \label{fig:bladder_perturbation}
\end{figure}

Figure~\ref{fig:bladder_perturbation} shows the estimated perturbation functions for the three models. The Bernstein estimate displays less stable finite-sample behaviour near the upper boundary, which may make the asymptotic linear regime difficult to identify empirically and may contribute to unstable tail-index estimation.

Hence, similar in-sample fits can lead to markedly different long-term extrapolations, highlighting the sensitivity of tail inference to the specification of the perturbation model. Given the primary emphasis on tail-index estimation, together with the censoring level observed in this dataset, the simulation results suggest that the Beta specification may provide the most stable estimate of $\xi$ in this setting. However, care is still required in selecting the appropriate model, as tail extrapolation remains highly sensitive to modelling assumptions.

\subsection{Heart failure data from INCOR--FMUSP}

As a second real-data application, we consider a heart failure dataset obtained from the Heart Institute of the University of São Paulo Medical School (INCOR--FMUSP), Brazil. The database comprises patients diagnosed with heart failure who entered follow-up between July 2003 and March 2014 and were subsequently monitored through regular clinical visits. The event of interest is all-cause mortality.

For each patient, the observed survival time was defined as the elapsed time between study entry and either the date of death or the date of the last recorded follow-up, whichever occurred first. Survival times are measured in months. Individuals who remained alive at the end of the study period or who were lost to follow-up were treated as right-censored observations.

After data cleaning and consistency checks, the final dataset consisted of $n=2,982$ patients, of whom 1,286 experienced the event of interest and 1,696 were censored, corresponding to a censoring proportion of approximately 57\%.

The same modelling strategy adopted in the previous application was employed. The parametric EGP model and the two semi-parametric EGP specifications based on Bernstein polynomials and histograms were fitted to the data. For the semi-parametric models, the complexity parameter was fixed at $10$. In addition, the {cure rate} model was included as a benchmark for comparison. Model performance was assessed through comparison with the Kaplan--Meier estimator and by examining the resulting tail behaviour and long-term survival extrapolations.

\begin{figure}[H]
\centering
\includegraphics[height=5.5cm]{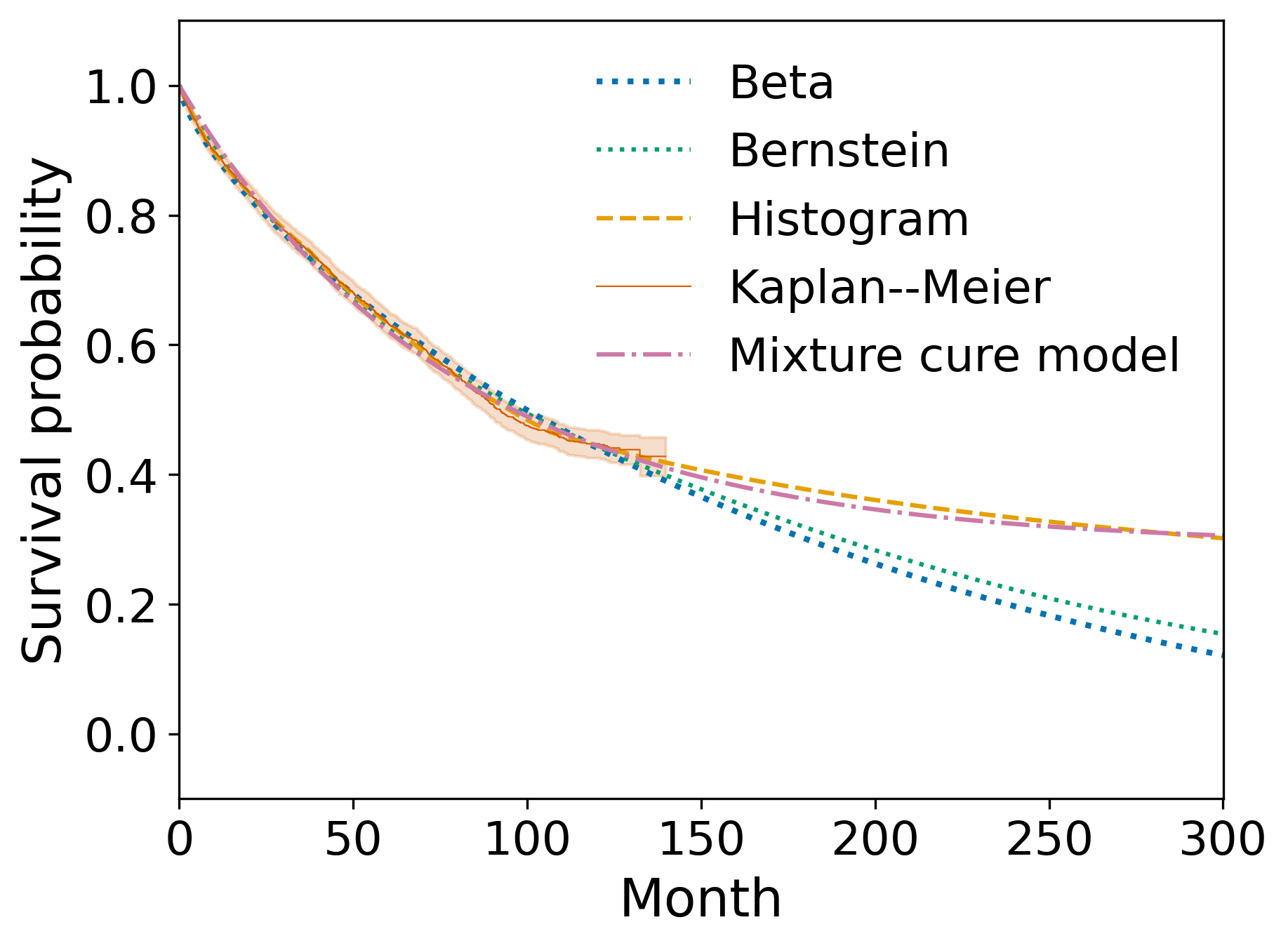}
\caption{Estimated survival functions for the INCOR heart failure dataset}
\label{fig:incor_survival}
\end{figure}

Figure~\ref{fig:incor_survival} displays the Kaplan--Meier estimator together with the fitted models. All specifications provide a satisfactory fit within the observed follow-up period, closely tracking the nonparametric estimate and remaining within its confidence bands. As in the bladder cancer application, the fitted curves are nearly indistinguishable over the data-supported region but exhibit markedly different long-term behaviour.

\begin{table}[htbp]
\centering
\caption{Estimated extreme value index $\xi$ and scale parameter for the three EGP  specifications}
\label{tab:xi_scale_endpoint}
\begin{tabular}{lcc}
\hline
EGP specification & $\hat{\xi}$ & $\hat{\sigma}$ \\
\hline
Beta       & $-0.3736$ & $229.2235$ \\
Bernstein  & $0.2449$  & $62.9014$  \\
Histogram  & $1.9734$  & $72.0589$  \\
\hline
\end{tabular}
\end{table}

The estimated extreme value indices for the Beta, Bernstein, and histogram models are $-0.37$, $0.25$, and $1.97$, respectively (see Table~\ref{tab:xi_scale_endpoint}). These markedly different estimates produce correspondingly different tail extrapolations despite the similarity of the fitted survival curves within the range of the observed data. According to the simulation results presented in the previous section, the combination of sample size and censoring level observed here would generally favour the Beta specification. Nevertheless, it is noteworthy that the Beta model is the only specification yielding a negative estimate of the extreme value index, implying a finite upper endpoint for the survival distribution.

To better understand these differences, we examine the estimated perturbation functions associated with the three EGP specifications.

\begin{figure}[H]
\centering
\includegraphics[height=5.5cm]{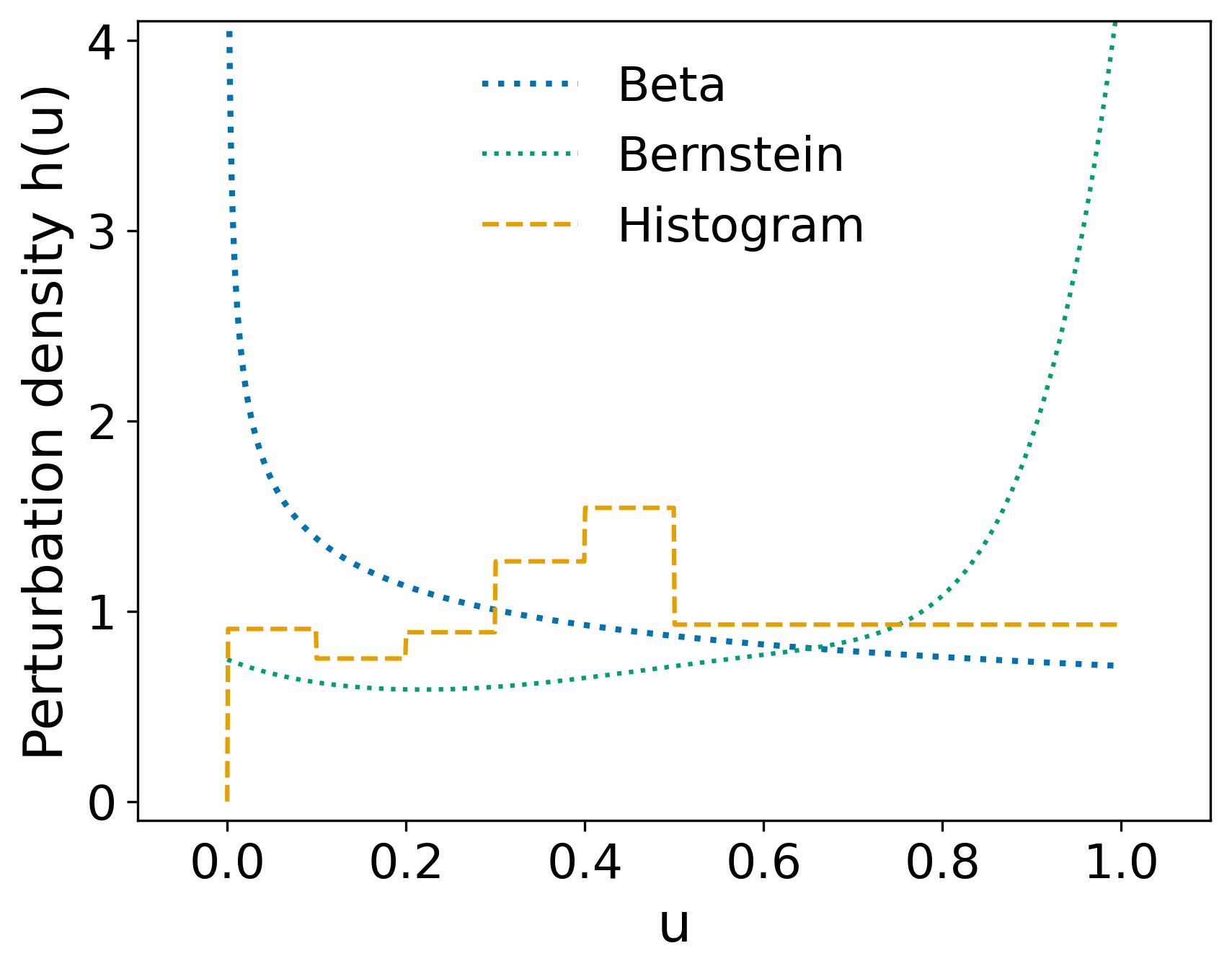}
\caption{Estimated perturbation densities for the INCOR heart failure dataset}
\label{fig:incor_perturbation}
\end{figure}

Figure~\ref{fig:incor_perturbation} presents the estimated perturbation functions for the three models. The overall pattern is remarkably similar to that observed in the bladder cancer dataset. In particular, the Bernstein specification exhibits less stable behaviour near the right boundary, whereas the Beta and histogram models display a more regular pattern in the neighbourhood of one. As discussed previously, deviations from the expected asymptotic linearity of the perturbation function may affect the preservation of the extreme value index and consequently influence tail estimation.

Overall, this application further illustrates that models providing virtually identical fits over the observed follow-up period may imply substantially different conclusions regarding long-term survival. The results reinforce the importance of carefully assessing the perturbation specification when employing EGP-based models for extrapolation beyond the range of the available data.

\section{Discussion and conclusion}

This work develops and investigates a class of EGP  models for right-censored survival data, with a particular focus on tail inference and long-term extrapolation. The main motivation is to combine the strengths of survival analysis, namely its principled handling of censoring, with the extrapolative capabilities of extreme value theory, while avoiding restrictive cure assumptions.

We proposed and compared three alternative specifications for the perturbation distribution within the EGP  framework: a parametric Beta model, a Bernstein polynomial estimator, and a histogram-based estimator. All models are embedded in a unified iterative estimation scheme adapted to right-censored data. This design choice ensures that differences in performance can be attributed to the perturbation structure rather than to differences in estimation procedures.

A comprehensive Monte Carlo study was conducted under a wide range of scenarios, varying tail heaviness, censoring intensity, sample size, and perturbation complexity. The results reveal a clear trade-off between flexibility and stability. The Beta specification, while highly structured, performs particularly well for the estimation of the tail index $\xi$ in terms of MAE, especially in moderate to large samples and under heavier-tailed regimes. In contrast, the histogram-based estimator provides the most consistent overall performance in terms of bias and is generally the most competitive specification for the scale parameter $\sigma$, particularly under low censoring. The Bernstein specification exhibits intermediate behaviour, but its relative performance depends more strongly on sample size and censoring, showing notable advantages in small samples and under high censoring.

Overall, no single model dominates uniformly across all criteria. Rather, the results indicate that model choice should be guided by the inferential objective and the characteristics of the data. Parametric structure can be advantageous for accurate tail index estimation, whereas nonparametric flexibility can provide more stable estimation of the scale parameter and reduce systematic bias across a broad range of scenarios. The bias results, in particular, highlight the robustness of the histogram specification across different tail indices, sample sizes, and censoring levels, while also showing that the Bernstein model can become preferable in specific challenging settings. Importantly, increasing flexibility through finer discretisation or higher polynomial degree does not necessarily improve performance and may even degrade estimation accuracy when sample sizes are limited. These findings emphasize that greater model flexibility should not be viewed as inherently beneficial, and that an appropriate balance between flexibility, stability, and the specific inferential target is crucial for reliable EGP estimation.

The two real-data applications further illustrate these findings. In both the bladder cancer and heart failure datasets, all EGP specifications provided very similar fits within the observed follow-up period and remained broadly consistent with the Kaplan--Meier estimator. Nevertheless, the models produced substantially different estimates of the extreme value index and consequently very different long-term extrapolations. These discrepancies were particularly evident in the heart failure application, where the estimated values of $\xi$ ranged from a negative value under the Beta specification to strongly positive values under the nonparametric alternatives. Despite these differences, the estimated perturbation functions exhibited similar patterns across the two datasets, reinforcing the connection between the behaviour of the perturbation near the upper boundary and the resulting tail inference.

Taken together, the simulation study and the empirical applications demonstrate that tail inference under the EGPD framework is highly sensitive to the specification of the perturbation distribution. While the proposed models offer considerable flexibility and are capable of accommodating a wide range of survival behaviours, the choice of perturbation model can have a profound impact on extrapolated survival probabilities beyond the observed data. Consequently, careful consideration of sample size, censoring level, and the primary inferential objective is essential when selecting among competing specifications.

Future research may focus on data-driven procedures for selecting the perturbation complexity, formal model selection criteria tailored to tail inference, and robustness analyses assessing the impact of perturbation misspecification on long-term survival predictions. Such developments would further strengthen the practical applicability of EGPD models as tools for survival analysis in the presence of censoring and limited follow-up.

\bibliographystyle{elsarticle-num.bst}
\bibliography{bibliografia}

\newpage
\pagenumbering{arabic}
\setcounter{table}{0}
\renewcommand{\thetable}{S\arabic{table}}
\section*{Supplementary simulation results}

This supplementary material reports the complete numerical results of the Monte Carlo study presented in the main text. The tables are organized by perturbation specification, namely the Beta, Bernstein, and histogram models within the Extended Generalized Pareto framework under right censoring.

For each configuration of sample size, tail index, censoring mechanism, and, where applicable, number of bins $m$, we report the empirical performance of the estimators for the Generalized Pareto distribution parameters $(\xi,\sigma)$. To accommodate the space constraints of the tables, we use the following notation: $cp$ denotes the censoring proportion, $\mathrm{Md}$ denotes the median of the estimates across Monte Carlo replications, $e$ denotes the mean absolute error (MAE), and $B$ denotes the bias. Thus, $\mathrm{Md}(\widehat{\sigma})$ and $\mathrm{Md}(\widehat{\xi})$ are the median estimates of $\sigma$ and $\xi$, respectively, while $e(\widehat{\sigma})$ and $e(\widehat{\xi})$ denote their corresponding MAEs, and $B(\widehat{\sigma})$ and $B(\widehat{\xi})$ their corresponding biases. The column $cp$ summarizes the empirical censoring proportion across replications.

These results complement the discussion in the main text by providing a detailed and systematic breakdown of the finite-sample behaviour of each estimator across all simulated scenarios.

Entries with $\ast$ correspond to cases where the numerical optimization procedure exhibited instability, preventing reliable convergence and resulting in unvalidated estimates.

\begin{longtable}[t]{cccccccccc}
\caption{\label{tab:tab:sim_beta}Simulation results for the Beta perturbation model}\\
\toprule
$n$ & $\xi$ & $\xi_c$ & $cp$ & $\mathrm{Md}$$(\widehat\sigma)$ & $e$$(\widehat\sigma)$ & $B$$(\widehat\sigma)$ & $\mathrm{Md}$$(\widehat\xi)$ & $e$$(\widehat\xi)$ & $B$$(\widehat\xi)$\\
\midrule
50 & 0.5 & -1 & 0.26 & 1.40 & 1.38 & 0.99 & -0.32 & 2.84 & -1.28\\
50 & 0.5 & -1 & 0.50 & 1.82 & 1.36 & 1.18 & -0.30 & 1.24 & -0.98\\
50 & 0.5 & -1 & 0.75 & 1.74 & 1.02 & 0.94 & 0.12 & 0.52 & -0.42\\
50 & 0.5 & 0 & 0.26 & 0.77 & 2.81 & 1.99 & -0.06 & 4.16 & -2.97\\
50 & 0.5 & 0 & 0.50 & 1.21 & 1.19 & 0.64 & 0.03 & 1.52 & -1.18\\

50 & 0.5 & 0 & 0.75 & 1.58 & 1.02 & 0.81 & 0.20 & 0.62 & -0.48\\
\bottomrule
\end{longtable}

\begin{longtable}[t]{ccccccccccc}
\caption{\label{tab:tab:sim_bernstein}Simulation results for the Bernstein perturbation model}\\
\toprule
$n$ & $\xi$ & $\xi_c$ & $m$ & $cp$ & $\mathrm{Md}$$(\widehat\sigma)$ & $e$$(\widehat\sigma)$ & $B$$(\widehat\sigma)$ & $\mathrm{Md}$$(\widehat\xi)$ & $e$$(\widehat\xi)$ & $B$$(\widehat\xi)$\\
\midrule
50 & 0.5 & -1 & 10 & 0.26 & 0.83 & 1.15 & 0.67 & -0.75 & 3.55 & -2.40\\
50 & 0.5 & -1 & 10 & 0.50 & 0.95 & 0.36 & 0.07 & 0.08 & 0.65 & -0.36\\
50 & 0.5 & -1 & 10 & 0.75 & 0.93 & 0.34 & 0.03 & 0.46 & 0.31 & -0.06\\
50 & 0.5 & -1 & 20 & 0.26 & 0.86 & 0.94 & 0.50 & -0.89 & 3.18 & -1.30\\
50 & 0.5 & -1 & 20 & 0.50 & 1.06 & 0.46 & 0.26 & -0.15 & 1.06 & -0.38\\

50 & 0.5 & -1 & 20 & 0.75 & 0.99 & 0.36 & 0.09 & 0.38 & 0.32 & -0.16\\
\bottomrule
\end{longtable}

\begin{longtable}[t]{ccccccccccc}
\caption{\label{tab:tab:sim_hist}Simulation results for the histogram perturbation model}\\
\toprule
$n$ & $\xi$ & $\xi_c$ & $m$ & $cp$ & $\mathrm{Md}$$(\widehat\sigma)$ & $e$$(\widehat\sigma)$ & $B$$(\widehat\sigma)$ & $\mathrm{Md}$$(\widehat\xi)$ & $e$$(\widehat\xi)$ & $B$$(\widehat\xi)$\\
\midrule
50 & 0.5 & -1 & 10 & 0.26 & 0.96 & 0.50 & 0.13 & 0.41 & 2.08 & 0.23\\
50 & 0.5 & -1 & 10 & 0.50 & 1.02 & 0.42 & 0.17 & 0.33 & 0.83 & -0.10\\
50 & 0.5 & -1 & 10 & 0.75 & 1.02 & 0.35 & 0.13 & 0.44 & 0.41 & -0.05\\
50 & 0.5 & -1 & 20 & 0.26 & 0.96 & 0.55 & 0.16 & 0.25 & 2.27 & 0.02\\
50 & 0.5 & -1 & 20 & 0.50 & 1.01 & 0.50 & 0.24 & 0.26 & 0.94 & -0.20\\

50 & 0.5 & -1 & 20 & 0.75 & 1.03 & 0.47 & 0.26 & 0.40 & 0.47 & -0.11\\
\bottomrule
\end{longtable}

\end{document}